\documentclass[aps,twocolumn]{revtex4}
\usepackage{epsfig}
\newcommand{\be}{\begin{equation}}
\newcommand{\ee}{\end{equation}}
\newcommand{\bea}{\begin{eqnarray}}
\newcommand{\eea}{\end{eqnarray}}
\newcommand{\ba}{\begin{array}}
\newcommand{\ea}{\end{array}}

\begin{document}

\title{The coil-globule transition of confined polymers} 
\author{Hsiao-Ping Hsu and Peter Grassberger}
\affiliation{John-von-Neumann Institute for Computing, Forschungszentrum
J\"ulich, D-52425 J\"ulich, Germany}
                                                                                
\date{\today}
\begin{abstract}
We study long polymer chains in a poor solvent, confined to the space between
two parallel hard walls. The walls are energetically neutral and pose only 
a geometric constraint which changes the properties of the coil-globule 
(or ``$\theta$-") transition. We find that the $\theta$ temperature increases 
monotonically with the width $D$ between the walls, in contrast to recent
claims in the literature. Put in a wider context, the problem can be seen 
as a dimensional cross over in a tricritical point of a $\phi^4$ model. 
We roughly verify the main scaling properties expected for such a phenomenon, 
but we find also somewhat unexpected very long transients before the 
asymptotic scaling regions are reached. In particular, instead of the  
expected scaling $R\sim N^{4/7}$ exactly at the ($D$-dependent) theta point
we found that $R$ increases less fast than $N^{1/2}$, even for extremely long 
chains.
\end{abstract}

\maketitle

\section{Introduction}

Thin films and quasi low dimensional systems are of obvious technological 
interest, in areas ranging from electronics to anti-corrosion coatings. Due 
to this, also critical phenomena in systems which are finite in one direction 
but infinite in all other ones have been studied in quite some detail
\cite{krech-dietrich,krech}. As the thickness of the film decreases, the properties cross 
over from bulk (3d) behaviour to surface (2d). Actually, in the true 
thermodynamic limit the system should show the scaling typical for 2d
critical phenomena for all finite thicknesses. But since the behaviour 
must, for finite systems, resemble those of 3d critical systems when the 
thickness is larger than the lateral size, the amplitudes in the thermodynamic
must show some special scaling. All this is known as {\it dimensional 
cross over}. The forces exerted by the critical fluctuations on the walls are 
known as Casimir forces \cite{krech}. The phenomenon exists also for tricritical 
points, although it has been studied much less for them.

Long flexible polymers in very diluted solutions can be described as the 
limit of the $\phi^4$ $O(n)$ vector model in the limit $n\to 0$ 
\cite{deGennes}. The coil-globule (``theta") transition happening as the 
solvent quality becomes worse (typically as temperature is lowered) is 
in this framework described as a tricritical point. 
Therefore, the 
problem of a polymer confined within the gap between two parallel plane 
walls is formally described as a dimensional cross over, either at a 
normal critical point (athermal polymers) or at a tricritical point 
(theta polymers), and the forces exerted by such a polymer are analogous 
to (critical/tricritical) Casimir forces.

But polymers have some special features which find no close analogy in $O(n)$ 
models with $n>0$. One of them is the fact that the volume occupied by an 
athermal polymer of fixed chain length $N$ and confined within two athermal
walls with distance $D$ does not depend monotonously on $D$ \cite{vliet}.
If $D$ is larger than the Flory radius $R_F\propto N^\nu$ (with $\nu\approx 
0.5876$), then the main effect of the confining walls is to reduce the 
size perpendicular to the walls, and thus both the gyration and the end-to-end
radius shrink with decreasing $D$. But when the polymer is strongly compressed
($D\ll R_F$), then the main effect is the lateral swelling due to the 
increased excluded volume interaction. In this case, $R^2 = R_\perp^2+R_\|^2$
and the occupied volume $R_\perp \times R_\|^2$ both increase
when $D$ is decreased further.   

This observation was the basis for a recent claim \cite{kumar} that the 
theta temperature $T_\theta$ of a polymer in a poor solvent between two 
athermal walls should also be non-monotonic in $D$. The argument is 
essentially that the monomer density controls the number of monomer-monomer
contacts, and thus also $T_\theta$: The higher is the density, the more 
effective will be the attractive monomer-monomer interaction, and the higher 
will be $T_\theta$. Therefore, as $D$ is increased, $T_\theta$ should start at
its 2d value, increase, go through a maximum, and finally decrease in order 
to reach its 3d value when $D\to\infty$. This predicted behaviour was then 
supported by exact enumeration studies of short chains ($N\leq 20$) on the 
simple cubic lattice (see Fig.~6 below).

In the present paper we show by means of rather extensive Monte Carlo 
simulations that this is not correct, and that $T_\theta$ increases 
monotonically with $D$. This is essentially what one would have expected 
from a dimensional cross over of a tricritical point. On the other hand, 
some of the details of this cross over are somewhat surprising. In particular, 
we find extremely long transients. At $T=T_\theta(D)$, the asymptotic scaling 
$R_\|\sim N^{\nu_{\theta,2}}$
with $\nu_{\theta,2} = 4/7$ being the Flory exponent for 2d theta polymers 
\cite{Duplantier,hegger} is not seen even when $R_F / D\approx 100$, 
although we have no reason to doubt that it will hold for $R_F / D\to \infty$.
The same is true for the free energy: Although we have no doubt that the 
scaling appropriate for 2d theta polymers will apply asymptotically at 
$T=T_\theta(D)$, it is not yet seen in the simulations. 

These simulations are done with the pruned-enriched Rosenbluth method 
(PERM) \cite{g97} which is ideally suited for this purpose. It allows to study
extremely long chains (for $D=60$ we went up to $N=600,000$) with very high
statistics, and it gives immediately very precise estimates of free energies.
Throughout the paper we shall model the polymers by self avoiding walks on 
the simple cubic lattice with attractive energy $-\epsilon$ between non-bonded
neighbouring monomers. Instead of quoting temperatures or values of $\epsilon$,
we shall describe thermal effects in terms of the Boltzmann factor
\be
   q = e^{\epsilon /k_bT}
\ee
per contact. The partition sum is therefore 
\be
   Z_N(q,D) = \sum_m C_{N,m}(D) q^m
\ee
where $C_{N,m}(D)$ is the number of walks with $N$ steps and $m$ monomer-monomer
contacts. The width $D$ is defined such that $D=1$ corresponds to the 
standard square lattice. For large $D$ we used hashing as described e.g.
in \cite{hg95} in order to minimize storage demands.

\section{Numerical Results}

In the following, we will denote by $q_\theta(D)$ the value of the Boltzmann
factor at the true quasi-2d theta point. The Boltzmann factor for 3-d theta 
polymers in the bulk is then $q^{(3)}_\theta = \lim_{D\to\infty}q_\theta(D)$, 
while the Boltzmann factor for strictly 2-d theta polymers on the square lattice 
is $q^{(2)}_\theta =q_\theta(D=1)$. The same notation is used for theta temperatures
and for growth constants (inverse critical fugacities). Growth constants at 
temperatures different from the collapse point will be denoted as $\mu(q,D)$, 
so that e.g. $\mu^{(3)}_\theta = \lim_{D\to\infty} \mu(q^{(3)}_\theta,D)$.

Exactly at the 3-d tricritical (theta) point, we can assume finite size scaling
ansatzes for the partition sum and for the rms. end-to-end distances both parallel
and perpendicular to the walls:

\be
   Z_N(q^{(3)}_\theta,D) \approx [\mu^{(3)}_\theta]^N \Phi(D/R_F(N))\;,
            \label{z0}
\ee
\bea
   R_{N,\|}(q^{(3)}_\theta,D) & = & \langle (x_N-x_0)^2+(y_N-y_0)^2\rangle \nonumber \\
                        & \approx & R_F(N) \Psi_\|(D/R_F(N))
            \label{r_para}
\eea
and
\be
   R_{N,\perp}(q_\theta,D) = \langle (z_N-z_0)^2 \rangle \approx R_F(N) \Psi_\perp(D/R_F(N))
            \label{r_perp} \;,
\ee
up to logarithmic corrections \cite{dupla,hg95,g97,hager-schaefer}. Here, $R_F(N)$
is the Flory radius (rms. end-to-end distance of chains in the bulk) which scales 
at the theta point like $N^{1/2}$, again up to logarithmic corrections. The 
scaling functions $\Phi(z), \Psi_\|(z)$, and $\Psi_\perp(z)$ are finite and non-zero
in the limit $z\to\infty$.
Finally, the growth constant $\mu^{(3)}_\theta$ is a non-universal constant 
which for the present model is $5.0479050\pm 0.0000005 + (q^{(3)}_\theta - 1.3087)\times 
1.616$ \cite{hegger,g97,unpub}.

If the theta temperature $T_\theta(D)$ increases with $1/D$, sufficiently long polymers at 
$q=q^{(3)}_\theta$ will be collapsed for finite $D$ (i.e., $R_N(q^{(3)}_\theta,D) \sim N^{1/2}$), 
while they will be swollen ($R_N(q^{(3)}_\theta,D) \sim N^{3/4}$) if 
$T_\theta(D)$ decreases with $1/D$. In Fig.~1 we plot $R^2_{N,\|}(q^{(3)}_\theta,D) / N^{3/2}$ 
against $N$, for various values of $D$. We see that all curves become horizontal 
for large $N$, i.e. all chains are swollen. This is in contradiction to the claim
of \cite{kumar}. The reason why the heuristics leading to this claim were wrong is
indeed quite clear: The location of the theta point is determined by polymers whose
Flory radius is much larger than $D$, while the anomaly noticed by van Vliet {\it 
et al.} \cite{vliet} concerns only polymers with $R_F \approx D$.
\begin{figure}
  \begin{center}
    \psfig{file=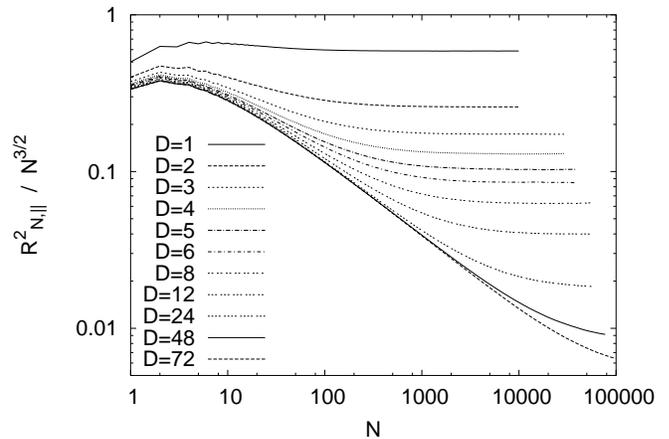,width=6.0cm,angle=270}
   \caption{Squared end-to-end distances parallel to the walls, divided by $N^{3/2}$,
     plotted against $N$. For self avoiding walks in 2d one has $R^2 \sim N^{3/2}$, 
     i.e. the walks are swollen coils if the lines become horizontal for large $N$.}
\label{r}
\end{center}
\end{figure}

From Eq.(\ref{r_para}) we see that $\Psi_\|(z) \sim z^{-1/2}$ for $z\to 0$, in order to obtain
$R_{N,\|}(q^{(3)}_\theta,D) \sim A(D) N^{3/4}$ with some amplitude $A(D)$. The $D$-dependence 
of the amplitude is then also fixed by Eq.(\ref{r_para}), 
\be
   A(D) \sim D^{-1/2} \;.     \label{ad0}
\ee
To check this, we plot in Fig.~2 the amplitudes obtained by fitting horizontal lines
to the large-$N$ data in Fig.~1. We see that the behaviour is roughly as predicted in 
Eq.(\ref{ad0}), but not quite. There are obviously substantial logarithmic corrections
(similarly large corrections are also seen in theta polymers in the bulk, see 
\cite{hg95,g97,hager-schaefer}). In the eyeball fit shown in Fig.~2 these are described
by a factor $\propto \ln(D)^{1/4}$, but this is just done to guide the eyes. A similarly
good fit would have been obtained with a pure power law $A(D) \sim D^{-0.54}$.

\begin{figure}
  \begin{center}
    \psfig{file=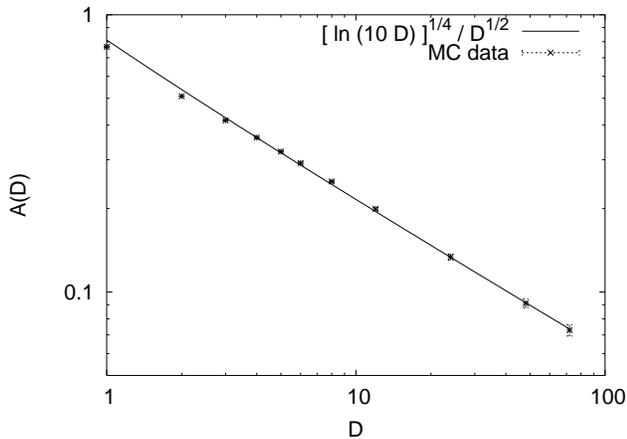,width=6.0cm,angle=270}
   \caption{Amplitudes for the end-to-end distance parallel to the walls (see Eq.(\ref{ad0}))
     plotted against $D$. The smooth line is an eyeball fit with a power law $\propto
     D^{-1/2}$ modified by a logarithmic term.} 
\label{ad}
\end{center}
\end{figure}

For all values of $D$, the partition sum for $N \gg D^{1/\nu_2}$ is also compatible 
with the scaling behaviour expected for 2-d SAWs ($\nu_2 = 3/4$ is the Flory exponent
in 2 dimensions),
\be
   Z_N(q^{(3)}_\theta,D) \sim Z_0(D) \mu(q^{(3)}_\theta,D)^N N^{\gamma_2-1}\;,
\ee
where $\gamma_2 = 43/32 = 1.34375$ and where $-k_BT^{(3)}_\theta \ln \mu(q^{(3)}_\theta,D)$ is 
the free energy per monomer in the thermodynamic limit. This is compatible with 
Eq.(~\ref{z0}), provided that
\bea
   {\mu(q^{(3)}_\theta,D)\over \mu^{(3)}_\theta} \approx 1-{a\over D^2}\;,    \label{muD} \\
    Z_0(D)\sim D^{2-2\gamma_2}\;, 
\eea
and 
\be
   \Phi(z) \sim \exp(-a/z^2) \qquad {\rm for}\quad z\to 0\;.
\ee
Eq.~(\ref{muD}) can also be derived from field theory 
\cite{krech-dietrich,krech,krech99,ritschel}, using De Gennes' mapping onto the 
$O(N=0)$ model. Numerical values for $\mu^{(3)}_\theta-\mu(q^{(3)}_\theta,D)$ are shown in Fig.~3. 
They are obviously in agreement with Eq.~(\ref{muD}) for large values of $D$, although
there are large corrections for finite $D$.

\begin{figure}
  \begin{center}
    \psfig{file=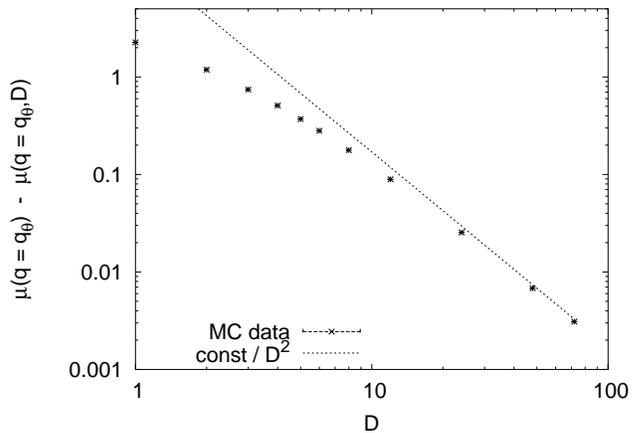,width=6.0cm,angle=270}
   \caption{Dependence of the effective connectivity constant $\mu(q,D)$ on $D$,
     for fixed $q = q^{(3)}_\theta$. The straight line gives the asymptotic behaviour 
     $\sim 1/D^2$, which is obviously modified by logarithmic corrections.
     Error bars on the MC data are much smaller than the symbol sizes.}
\label{mu_D}
\end{center}
\end{figure}

Up to now we have only discussed the behaviour at the theta point of the free 
($D=\infty$) polymer. Let us finally study the behaviour at and in the vicinity 
of the theta points for the quasi-2d systems at fixed finite $D$. According to
the above results, this corresponds, for any finite $D$, to temperatures where 
the free 3-d polymer would be collapsed. Asymptotically, for $N \to\infty$, the 
behaviour exactly at the tricritical point should be that for 2-d theta polymers,
\be
   Z_N(q_\theta(D),D) \sim [\mu(q_\theta(D),D)]^N N^{\gamma_{\theta,2}-1}
             \label{Z_2d_theta}
\ee
and
\be
   R_{N,\|}(q_\theta(D),D) \sim N^{\nu_{\theta,2}}\;.
            \label{r_2d_theta}
\ee
Here, $\nu_{\theta,2} = 4/7$ and $\gamma_{\theta,2}=8/7$ are the Flory and entropic 
exponents for 2-d theta polymers.

In Figs.~4 and 5 we show $R^2_{N,\|}(q,D)/N^{8/7}$ versus $N$, for several 
values of $q$ close to (tri-)criticality, and for $D=5$ (Fig.~4) resp. $D=60$
(Fig.~5). According to Eq.(\ref{r_2d_theta}), one of these curves in each 
graph should become horizontal for $N\to\infty$. Naively, one might expect 
this regime to set in when the (3-d) Flory radius is roughly equal to $D$, 
i.e. for $N\approx 25$ in Fig.~4 and for $N\approx 4000$ in Fig.~5. We do not see 
this, although our values of $N$ are much larger than these, by factors more 
than one hundred. Although we cannot pin down precisely $q_\theta(D)$ due to 
this, we definitely see that $R_N$ increases at the ($D$-dependent) theta point
-- and for numerically accessible values of $N$ -- slower than $\sqrt{N}$ (straight 
lines in Figs.~4 and 5). Thus instead of being swollen, as predicted by theory,
the polymers seem to be {\it collapsed} at the 3-d theta point. Similar results were 
found for all other values of $D$, and they are corroborated by the results obtained 
for $Z_N$. Also for $Z_N$, the asymptotic behaviour stated in Eq.(\ref{Z_2d_theta})
was not seen for any of the chains (data not shown).

\begin{figure}
  \begin{center}
    \psfig{file=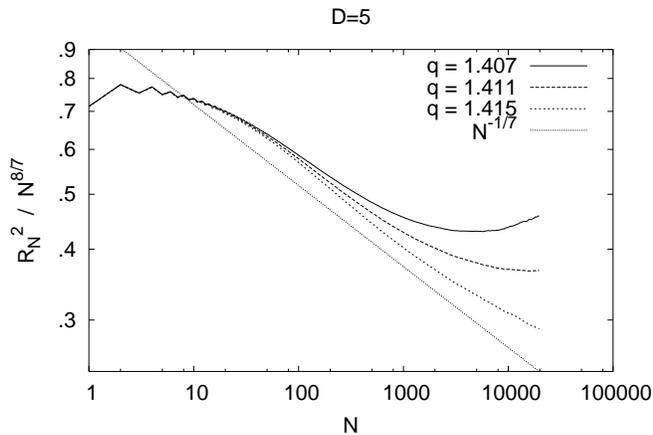,width=6.0cm,angle=270}
   \caption{Average squared end-to-end distances parallel to the walls for nearly tricritical 
     polymers in a slit of width $D=5$. More precisely, $R^2_N/N^{8/7}$ is plotted
     against $\ln N$, which would lead to a horizontal curve if Eq.(\ref{r_2d_theta})
     were satisfied for all $N$. The straight line indicates, in contrast, a 
     non-swollen behaviour $R^2_N\propto N$.}
\label{R5}
\end{center}
\end{figure}

\begin{figure}
  \begin{center}
    \psfig{file=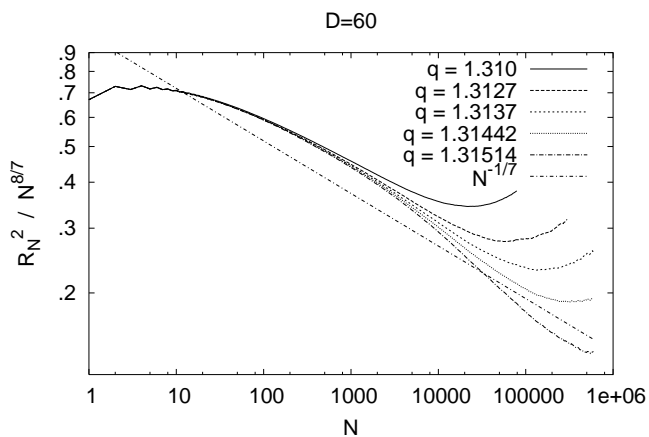,width=6.0cm,angle=270}
   \caption{Same as Fig.~\ref{R5}, but for $D=60$. Notice that the longest chains 
     in this case have $N=600,000$.}
\label{R60}
\end{center}
\end{figure}

While these results might look very surprising at first sight, it is indeed not 
too difficult to understand them heuristically. As we said, free 3-d chains would 
be slightly collapsed at the temperatures shown in Figs.~4 and 5. In a blob 
picture, these chains are therefore chains of blobs, each one of size $\approx D$,
and each representing a short 3-d polymer slightly below the theta point. It 
is well known \cite{hg95,g97,hager-schaefer} that the 3-d theta collapse of finite 
chains happens at an effective (``Boyle") temperature $T_\theta(N)$ which is lower
than the true $T^{(3)}_\theta$, and that concatenating such chains at $T_\theta(N)$
leads to a longer chain which is much more collapsed than its short constituents:
For temperatures slightly below $T^{(3)}_\theta$, there is a regime in $N$ where $d R_N / 
d N <0$, i.e. chains actually shrink when more monomers are attached to them! For 
the present problem this means that each blob might be swollen, but when two blobs 
are brought in contact, they might not repel each other (as in a swollen polymer), 
but they rather attract each other. Obviously this is what happens. For much longer 
chains than those we can simulate, this would finally stop: When too many blobs are 
penetrating each other, repulsion finally dominates again and the chains behave as
ordinary 2-d theta polymers.

\begin{figure}
  \begin{center}
    \psfig{file=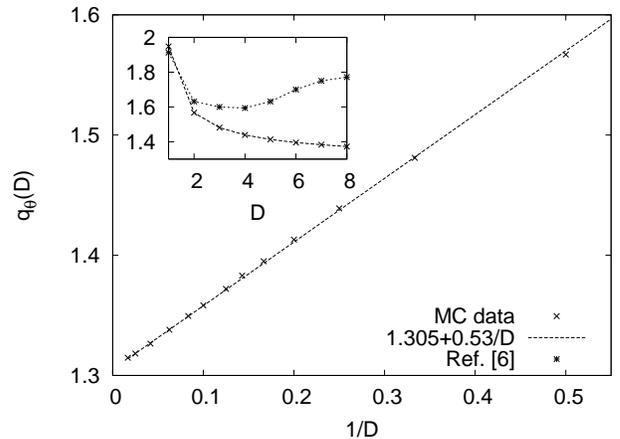,width=6.0cm,angle=270}
   \caption{Lower limit estimates of $q_\theta(D)$, plotted against $1/D$.
     The straight line is $q = 1.305+0.53/D$. The insert shows same of
these data plotted against $D$, together with the data from 
Ref.~[6] for comparison.}
\label{qc_D}
\end{center}
\end{figure}

According to this discussion we can give only lower bounds on 
$q_\theta(D)$. They are given by those values of $q$ for which the curves in
plots like Figs.~4 and 5 become horizontal. These bounds, which should however be not 
too far from the true values of $q_\theta(D)$, are shown in Fig.~6. We see 
an essentially linear increase with $1/D$. Extrapolating linearly to $1/D\to 0$ we 
find a value $q=1.305 \pm 0.001$ which is close to, but definitely smaller than, 
$q^{(3)}_\theta = 1.3087 \pm 0.0003$ \cite{g97}.
This confirms that the curves in Figs.~4 and 5 which become horizontal at the 
largest values of $N$ are not yet the critical ones, and that the true asymptotic
behaviour is not yet seen in these figures.
  
\section{Discussion}

On the one hand we have shown that some recent claims about re-entrant behaviour 
of polymer collapse in restricted geometries are wrong. The theta collapse of a 
polymer confined to the space between two parallel walls is very much as expected
from a tricritical behaviour in a film (quasi-2d) geometry. In particular, the 
collapse temperature is, for any distance $D$ between the walls, shifted to 
temperatures lower than the theta temperature of free 3-d polymers, and the 
decrease of the free energy with decreasing $D$ is as predicted from the theory
of tricritical Casimir effects.

But on the other hand, we found that the detailed behaviour at the true quasi-2d
theta collapse is -- for any chain length we could simulate, and presumably also 
for any chain length realistic in any foreseeable experiment -- rather different
from the predicted one. This is related to the fact that the upper critical 
dimension of theta collapse (as of any other tricritical phenomenon) is $d=3$. 
Therefore long chains are, at the collapse point, composed of blobs which
are essentially free random walks but which slightly attract each other.

In view of the analogy between theta collapse and other tricritical phenomena, 
it is of interest to speculate whether similar anomalies should be expected also 
for the latter. For tricritical Ising or Potts models one might then expect 
that the effective correlation length exponent is not given by the true 
tricritical exponent for $d=2$ (which is larger than $1/2$ for all these models), 
but has a value $<1/2$. To our knowledge this has neither been predicted so far,
nor has it been seen in simulations or in real experiments.

Acknowledgements: We thank Walter Nadler for very useful discussions, Michael
Krech for correspondence, and Sanjay Kumar for getting us interested in this 
problem.

\end{document}